\documentclass[aps,prl,preprint,groupedaddress,showpacs]{revtex4-1}

\usepackage{graphicx}
\usepackage{amsfonts}

\begin{document}

\title{Truncated L\'evy Flights and Weak Ergodicity Breaking in the Hamiltonian Mean Field Model}
\author{A.~Figueiredo}
\affiliation{Instituto de F\'\i{}sica and International Center for Condensed Matter Physics\\ Universidade de
Bras\'\i{}lia, CP: 04455, 70919-970 - Bras\'\i{}lia, Brazil}
\author{Z.~T.~Oliveira Jr}
\affiliation{Departamento de Ci\^encias Exatas e Tecnol\'ogicas,\\
Universidade Estadual de Santa Cruz,\\ CEP 45662-900 - Brazil}
\affiliation{International Center for Condensed Matter Physics\\ Universidade de
Bras\'\i{}lia, CP: 04455, 70919-970 - Bras\'\i{}lia, Brazil}
\author{T.~M.~Rocha Filho}
\affiliation{Instituto de F\'\i{}sica and International Center for Condensed Matter Physics\\ Universidade de
Bras\'\i{}lia, CP: 04455, 70919-970 - Bras\'\i{}lia, Brazil}
\author{R.~Matsushita}
\affiliation{Departamento de Estat\'\i stica\\Universidade de
Bras\'\i{}lia, CP: 04455, 70919-970 - Bras\'\i{}lia, Brazil}
\author{M.~A.~Amato}
\affiliation{Instituto de F\'\i{}sica and International Center for Condensed Matter Physics\\ Universidade de
Bras\'\i{}lia, CP: 04455, 70919-970 - Bras\'\i{}lia, Brazil}
\begin{abstract}
The dynamics of the Hamiltonian mean field model is studied in the context
of continuous time random walks. We show that the sojourn times in cells in the momentum
space are well described by a one-sided truncated L\'evy distribution. Consequently the system is non-ergodic
for long observation times that diverge with the number of particles.
Ergodicity is attained only after very long times both at thermodynamic equilibrium and at
quasi-stationary out of equilibrium states.
\end{abstract}

\maketitle

Ergodicity is a fundamental concept in statistical mechanics~\cite{birkhoff,kinchin} and essentially states that ensemble 
and time averages
are equal over a single trajectory of the system or, equivalently, that the sojourn time on a given region
is proportional to the ensemble measure. 
For particles subjected to an external potential, ergodicity implies that time averages over a
single particle trajectory is equal to the average over many particles at a fixed time. The latter case has recently been observed
experimentally for the diffusion of molecules on a nanostructured porous glass~\cite{feil}.
Strong ergodicity breaking occurs if some region
in the phase space is not accessible by the system trajectory.
On the other hand weakly non-ergodic behavior corresponds to a situation
where every state can be reached but the occupation statistics is not equal to the ensemble measure~\cite{bouchaud}.
The statistical mechanics of a system with weak ergodicity breaking, henceforth called weakly non-ergodic,
in the context of Continuous Time Random Walks (CTRW)
was addressed by Rebenshtok and Barkai~\cite{weba}.
The system can be in $M$ different states, such that a given observable ${\cal O}$ admits the respective $M$ values ${\cal O}_k$
for $k=1,\ldots,M$. The time average of this observable is then
\begin{equation}
\overline{\cal O}=\frac{1}{t_{\rm tot}}\sum_{k=1}^M t_k^{(r)}{\cal O}_k,\hspace{5mm}t_{\rm tot}=\sum_{i=k}^Mt_k^{(r)},
\label{eq1}
\end{equation}
where $t_k^{(r)}$ is the residence time, i.~e.\ the total time spent by the system in state $k$ and $t_{\rm tot}$ the total observation time.
The sojourn time $t_k^{(j)}$ is the time spent in state $k$ during the $j$-th visitation, and therefore $t_k^{(r)}=\sum_j t_k^{(j)}$.
Owing to L\'evy generalized central limit theorem,
the probability distribution of residence time $t_k^{(r)}$ can be described by a one-sided L\'evy distribution $f^{(\alpha)}_k(t)$ with
characteristic function~\cite{weba,Samorodnitsky}:
\begin{equation}
\psi_k(z)=\exp\left\{-\gamma_k|z|^\alpha\left[1-i\tan\left(\pi\alpha/2\right){z}/{|z|}\right]\right\},
\label{eq2}
\end{equation}
for $0<\alpha\leq2$ and $\alpha\neq1$, $\alpha=2$  corresponding to the Gaussian distribution,
and $\gamma_k$ is a constant scale factor~\cite{Samorodnitsky}. 
An important feature is that the moments $\langle t^\mu\rangle$ of the L\'evy distribution diverge for $\mu>\alpha$.
The distribution of the possible different values of the time averages $\overline{\cal O}$ is given by~\cite{weba}:
\begin{equation}
F^{(\alpha)}(\overline{\cal O})=-\frac{1}{\pi}\lim_{\epsilon\rightarrow0}{\rm Im}\left\{\left[
\sum_{k=1}^M\rho_k(\overline{\cal O}-{\cal O}_k+i\epsilon)^{\alpha/2-1}
\right]\left[{\sum_{k=1}^M\rho_k(\overline{\cal O}-{\cal O}_k+i\epsilon)^{\alpha/2}}\right]^{-1}\right\}.
\label{eq3}
\end{equation}
In the limit $\alpha\rightarrow2$ Eq.~(\ref{eq3}) reduces to $F^{(2)}(\overline{\cal O})=\delta(\overline{\cal O}-\langle{\cal O}\rangle)$,
where $\langle{\cal O}\rangle=\sum_k \rho_k{\cal O}_k$, and the coefficients $\rho_k$ are stationary probabilities.
As far as $\alpha\neq 2$ and consequently $F^{(\alpha)}\neq\delta(\overline{\cal O}-\langle{\cal O}\rangle)$ the system is weakly non-ergodic.
The extension of this approach for cells with different occupation statistics is given in Ref.~\cite{venegeroles}.
Examples of weakly non-ergodic systems include laser cooling of trapped atoms~\cite{saubamea}, diffusion of lipid granules in living cells~\cite{jeon},
blinking quantum dots~\cite{brokmann,margolin} and glass dynamics~\cite{bouchaud}.
In this paper we show that a classical Hamiltonian system with long-range interaction also displays a weakly non-ergodic behavior.

Systems with long-range interactions are characterized by a pair-interaction potential decaying asymptotically as $r^{-a}$, $a<D$
with $D$ the spatial dimension~\cite{dauxois}.
These systems have drawn some attention in the last two decades~\cite{physrep,proc1,proc2,proc3}, and have unusual properties
such as anomalous diffusion, aging, negative heat capacity at equilibrium, non-Gaussian quasi-stationary states and a relaxation time to equilibrium
diverging with the number $N$ of particles. This last feature seems ubiquitous in such systems and is related to
a very long time required to attain ergodicity, as we explicitly show bellow for one model system as a preliminary
step of a thorough investigation of general classes of long-range interacting systems.
Despite great interest, their dynamics is not completely understood due to inherent difficulties
in both analytic and numerical approaches. 
A few toy models were introduced in the literature in order
to simplify the understanding of their intricate behavior. Among them we mention the Hamiltonian Mean Field (HMF) model~\cite{hmforig},
which has become a sort of ground test for many numerical and analytic studies, and defined by the Hamiltonian:
\begin{equation}
H(p,\theta)=\frac{1}{2}\sum_{i=1}^N p_i^2+\frac{1}{2N}\sum_{i,j=1}^N\left[1-\cos(\theta_i-\theta_j)\right].
\label{eq4}
\end{equation}
This is a solvable system at equilibrium and can be interpreted as consisting of $N$ classical
rotors globally coupled with unit moment of inertia~\cite{hmforig,metodo}.
It has a second order phase transition from a spatially homogeneous to an inhomogeneous phase, and
a rich structure of non-equilibrium phase transitions~\cite{noneq}.
In the limit $N\rightarrow\infty$ it is described by the mean field Vlasov kinetic equation~\cite{braun}.
Thence we only have to consider the dynamics of a single particle evolving in the mean field
of the remaining particles. Quasi-stationary states thus correspond to the infinite number of stable stationary
states of the Vlasov equation. For finite $N$, small collisional corrections must be considered resulting in a secular evolution.

In the present letter the ergodic properties of the HMF model are studied by considering its dynamics as a CTRW
by dividing the momentum space in finite width cells, with each cell being considered as one possible state of the
system.
We observe a very slow convergence of $\alpha$ in Eq.~(\ref{eq3})
to the Gaussian value $\alpha=2$ for the sojourn times in the cells.
This is related to the very long relaxation time to thermodynamic equilibrium in
long-range interacting systems. Indeed, as discussed in
Ref.~\cite{eplnos}, the time of observation required for the system to attain ergodicity is of same order of magnitude as the relaxation time to
equilibrium. This kind of sluggish convergence is also associated to a truncated power
law tail in the sojourn times distribution~\cite{mantegna,nano}.
This implies that for shorter observation times the system can be considered as weakly non-ergodic,
ergodicity being only attained asymptotically.
One possible source of ergodicity breaking in long-range interacting systems, but not the only one, is a parametric resonance
of particle motion with mean-field oscillations~\cite{benetti}.
\begin{figure}[ptb]
\begin{center}
\scalebox{0.3}{{\includegraphics{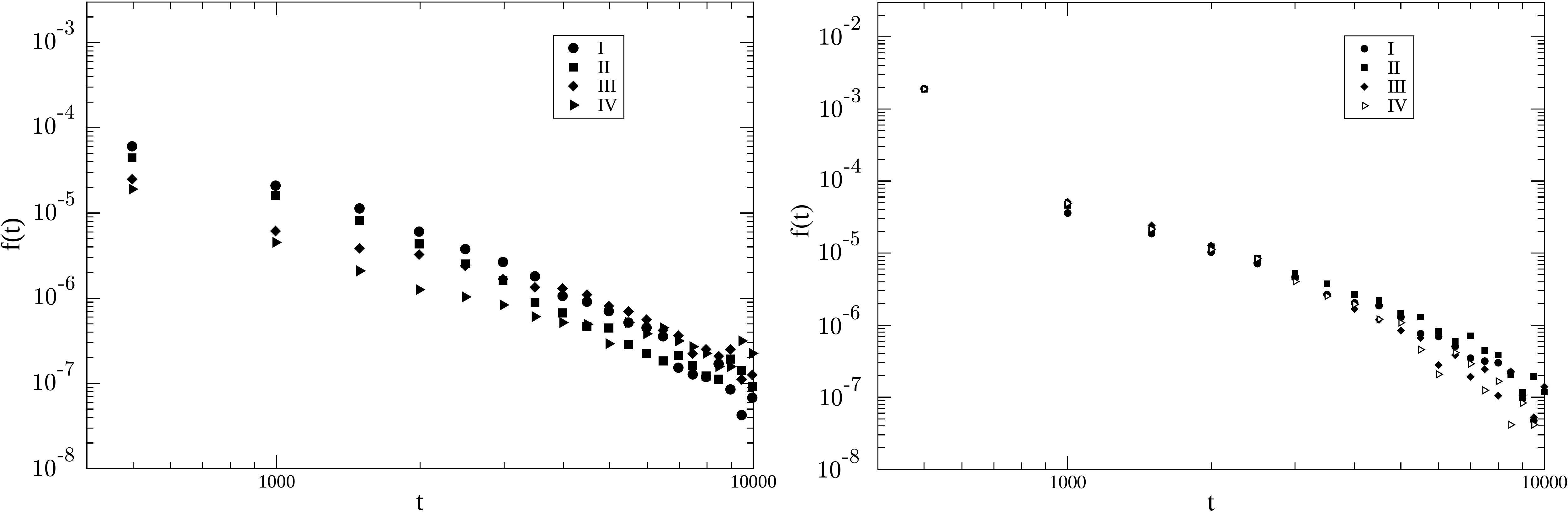}}}
\end{center}
\caption{Left Panel: Distribution $f(t)$ of sojourn times at the thermodynamic equilibrium state for
cells in momentum space in the intervals (I) $[0.0,0.4]$ (II) $[0.4,0.8]$,
(III) $[0.8,1.2]$ and (IV) $[1.2,1.6]$, $N=100,000$ and total simulation time $t_{\rm tot}=10^7$.
Right Panel: Same as left panel but for the waterbag state.}
\label{fig1}
\end{figure}

The dynamics of the HMF model is studied by performing molecular dynamics simulations for the Hamiltonian in Eq.~(\ref{eq4}).
The momentum space is divided in cells of width $\Delta p$, and each cell is then taken as a different discrete state.
We consider here two types of statistically stationary states, both spatially homogeneous:
(i) thermodynamic equilibrium, and (ii) a stable non-Gaussian (quasi) stationary state
with a waterbag one-particle constant momentum distribution in an interval $\rho(p)=1/2p_0$ if $-p_0<p<p_0$ and zero otherwise.
Spatially homogeneous distributions imply a vanishing force in the mean-field limit, but with small
fluctuations due to collisional corrections for finite $N$ causing small cumulative changes in the momentum distribution.
Simulations are performed using a symplectic solver for the Hamiltonian equations implemented in a parallel code in
graphic processing units~\cite{eu2}.
The temperature for the equilibrium case is $T=0.8$ and for the non-equilibrium case we chose
$p_0\approx2.68$ and total energy per particle $e=0.8$.

\begin{figure}[ptb]
\begin{center}
\scalebox{0.3}{{\includegraphics{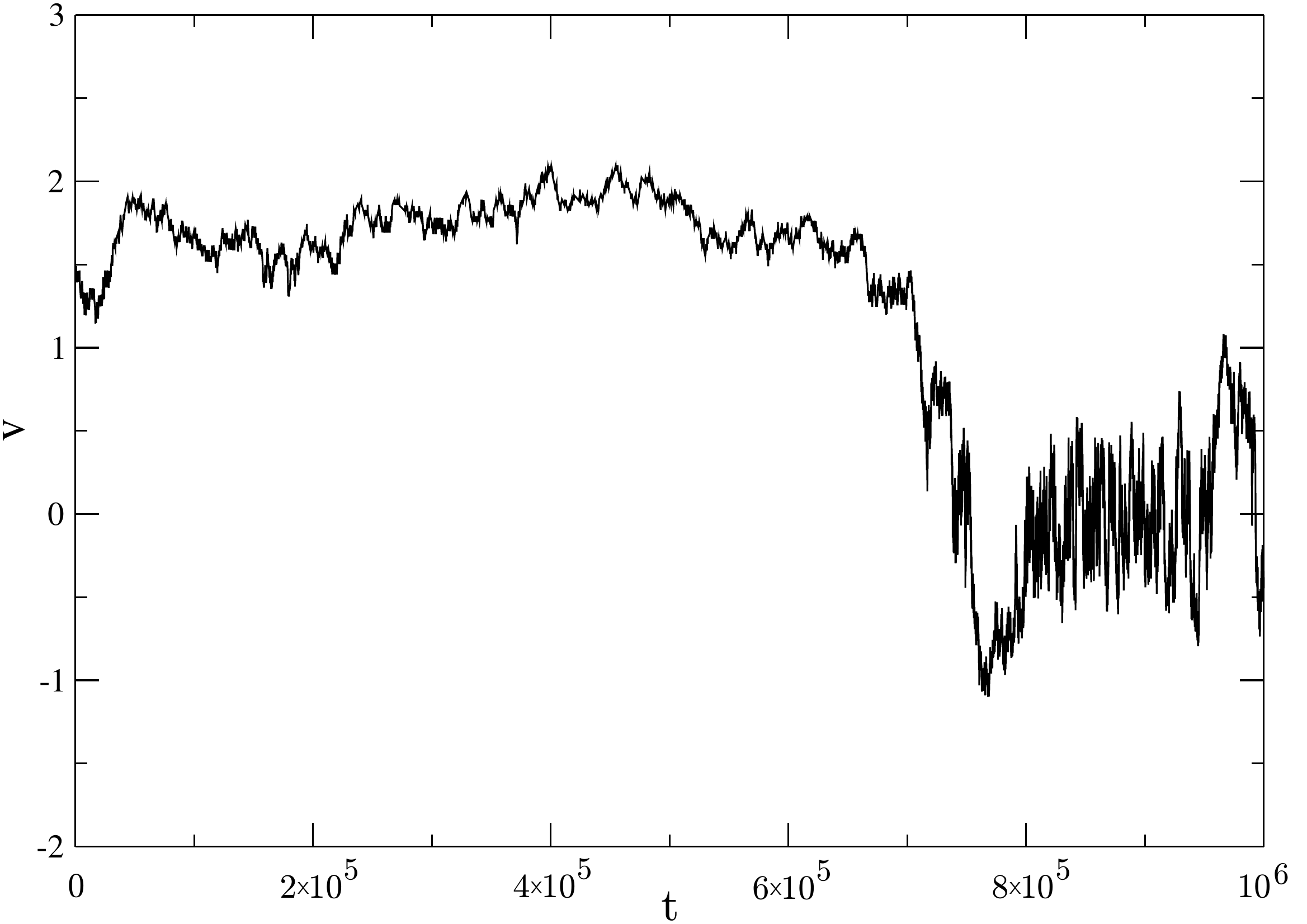}}}
\end{center}
\caption{Velocity of a particle as a function of time for the Hamiltonian in Eq.~(\ref{eq3}) with $N=50,000$
and total simulation time $t_f=10^6$ for the thermodynamic equilibrium distribution.}
\label{fig2}
\end{figure}

Figure~\ref{fig1} shows that the statistics of sojourn times for different cells of width $\Delta p=0.4$
are equal up to the noise level.
The time evolution of the velocity of a single particle is shown in Fig.~\ref{fig2}
illustrating clearly the trapping of the velocity in regions of the momentum space with rapid movements between different traps.
This is due to the system being in a homogeneous state and therefore the interactions on each particle correspond to
collisions whose effects diminish with increasing $N$, in such a way that either
a rare strong collision or the cumulative effects of weaker collisions
are required to extract the particle from a given cell.
As the statistics of sojourn times are approximately the same for different cells, sojourn times are taken from all cells thence allowing
a better statistical accuracy.  Indeed,
the sequence of sojourn times
is obtained by considering successive time intervals between the entrance and exit of any particle in any individual cell (or state).

The generalized central limit theorem lead us to expect sojourn time statistics 
$t^{(j)}=t_k^{(j)}$ for cells of width $\Delta p$ close to a one-sided L\'evy distribution (with 
positive support). We define the empirical normalized frequency of sojourns as \begin{equation} 
\label{sojourn} g(t)=\frac{1}{J}\sum_{j=1}^J\delta\left(t-t^{(j)}\right); \end{equation} where 
$t^{(j)}$ ($j=1,\ldots,J$) are the sojourn times obtained from the simulation. We drop the 
index $k$ associated to each cell (or state) because the sojourn time statistics are almost 
identical for cells with same width (see Figure~\ref{fig1}). We may expect a truncated one sided 
L\'evy distribution as a good approximation to describe $g(t)$. A truncated one sided L\'evy 
distribution is defined as follows: \begin{equation} \label{ann0} 
\tilde{f}^{(\alpha)}(t)=\left\{\begin{array}{ll} c f^{(\alpha)}(t) & 0<t<L \\ 0 & 
\mbox{otherwise} \end{array}\right. \end{equation} where $c$ is a normalization constant, 
$f^{(\alpha)}(t)$ is a L\'evy asymmetric distribution with characteristic function given by 
equation (\ref{eq2}) and $L$ is the truncature length. In this case, as the sojourn time 
statistics are taken as independent of the specific cell considered, we have $\gamma_k=\gamma$ 
for all $k$. 

\begin{table} \begin{center} \begin{tabular}{ccc | ccc} \hline\hline $N$ & $\alpha_{\rm eq}$ & $\alpha_{\rm wb}$ & $N$ & $\alpha_{\rm eq}$ & $\alpha_{\rm 
wb}$\\ \hline\hline 10,000 & \hspace{3mm} 0.67 \hspace{3mm} & 1.00\hspace{3mm} &\hspace{3mm} 80,000 \hspace{3mm} & 0.57 \hspace{3mm} & 0.66\\
 20,000  &    0.62    &     0.82  &  90,000  &    0.57    &     0.65\\
 30,000  &    0.60    &     0.74  & 100,000  &    0.57    &     0.64\\
 40,000  &    0.59    &     0.70  & 200,000  &    0.55    &     0.61\\
 50,000  &    0.58    &     0.69  & 300,000  &    0.55    &     0.61\\
 60,000  &    0.58    &     0.67  & 400,000  &    0.55    &     0.62\\
 70,000  &    0.57    &     0.66  & 500,000  &    0.56    &     0.63\\
\hline
\end{tabular}
\end{center}
\caption{Values of $\alpha$ in Eq.~(\ref{eq2}) for the thermodynamic equilibrium ($\alpha_{\rm eq}$) and
waterbag distribution ($\alpha_{\rm wb}$).}
\label{tab1}
\end{table}

Truncated L\'evy distributions were introduced by Mantegna and Stanley in a different context~\cite{mantegna}.
Such functions have finite moments, and therefore the sum of random variables
$X^{(n)}=\sum_{j=1}^n t^{(j)}$, with $t^{(j)}$ a random variable drawn from the truncated one sided L\'evy distribution,
converges to a Gaussian distribution.
Nevertheless this convergence is extremely slow and
the distribution of this sum remains close to a true stable L\'evy distribution for values of $n$ that can be very large,
displaying a typical scaling before a crossover to a Gaussian behavior for $n\gg n^\star$ where $n^\star$ denotes the crossover value for $n$ between
the L\'evy and Gaussian distributions $P(X^{(n)})$ for $X^{(n)}$. For all $n\ll n^\star$ the distributions $P(X^{(n)})$ collapse in the same
function under the scaling:
\begin{equation}
\tilde X^{(n)}= n^{-1/\alpha}\:X^{(n)}\:\: {\rm and}\:\: P(\tilde X^{(n)})= n^{1/\alpha}\:P(X^{(n)}).
\label{eq3e}
\end{equation}
It should be pointed out that as the cells with same width are almost statistically
identical, then $X^{(n)}$ can be taken as a random variable corresponding to the residence time $t^{(r)}_k$, and in view
of that, $P\left(X^{(n)}=t\right)$ is approximately a convolution of the L\'evy one sided $f^{(\alpha)}(t)$ for $t<L$.
\begin{figure}[ptb]
\begin{center}
\scalebox{0.75}{{\includegraphics{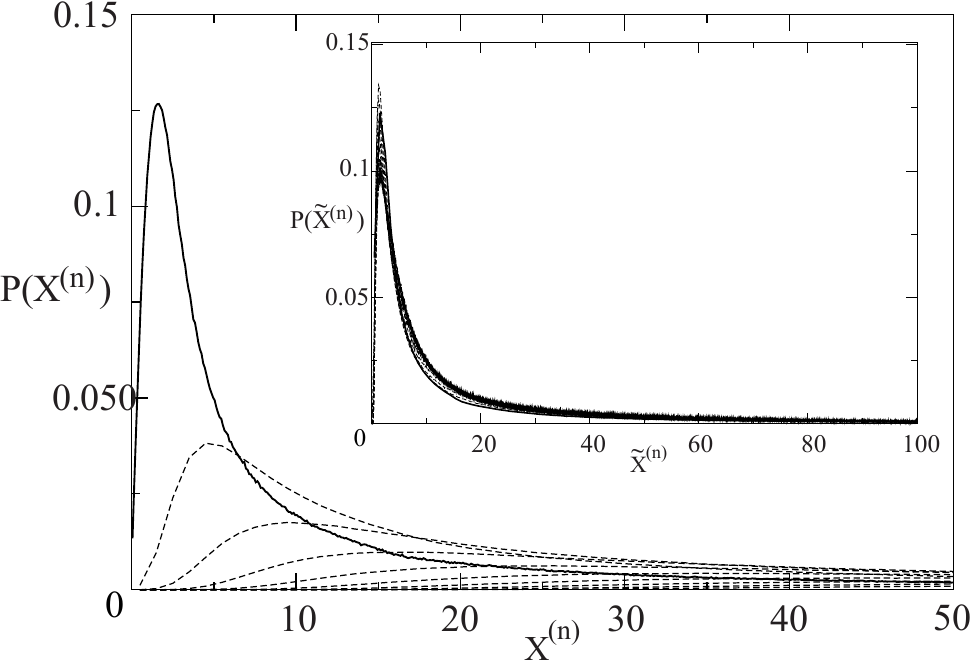}}}
\end{center}
\caption{Density distribution $P\left(X^{(n)}=t\right)$ for the sum $X^{(n)}$ of random variables with $n=1,3,5,7,9,11$ for the equilibrium
case and $N=200,000$ particles,
where higher values of $n$ has a smaller maximum. The continuous line corresponds to the distribution of sojourn times ($n=1$).
Inset: data collapse of the distribution functions by the scaling in Eq.~(\ref{eq3e}) with $\alpha=0.55$ as the
best estimate in table~\ref{tab1}.}
\label{fig5}
\end{figure}

The shape exponent $\alpha$ in Eq.~(\ref{eq2}) for the distribution of sojourn times are given in table~\ref{tab1} for the
equilibrium and non-equilibrium (waterbag) cases obtained using a maximum likelihood method~\cite{fauerverger,rao}.
The difference in the values of $\alpha$ for the equilibrium and waterbag cases with same $N$ are due to the
different particle velocities and variations
of the force field fluctuations around its zero average value (homogeneous state) for different energies.
Figure~\ref{fig5} shows the sojourn times distribution function for the distributions of summed variables
$X^{(n)}$ obtained by randomly summing $n$ sojourn times obtained from our simulation.
The inset shows the data collapse of these distribution when rescaled according
to Eq.~(\ref{eq3e}) as expected for a one-sided L\'evy distribution.
\begin{figure}[ptb]
\begin{center}
\scalebox{0.5}{\includegraphics{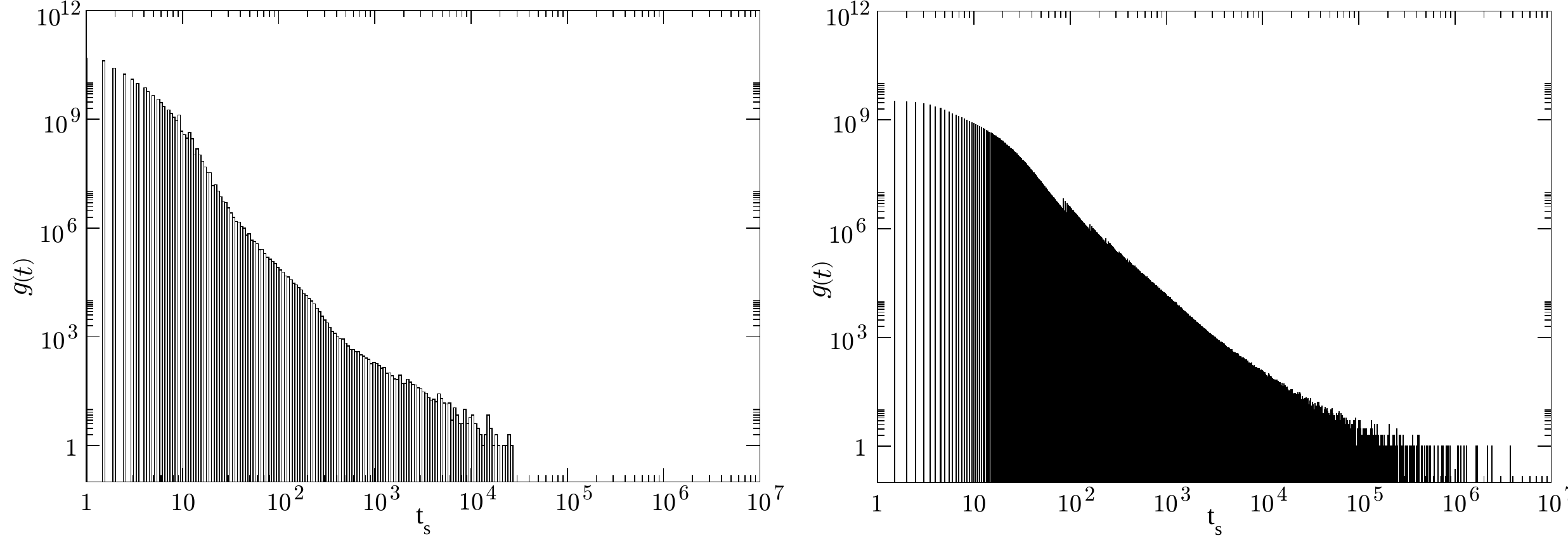}}
\end{center}
\caption{Left Panel: Histogram of the number of sojourn times for $N=50,000$ particles, $\Delta p=0.01$ and total integration time $t_f=10^7$.
Right Panel: The same but with $\Delta p=0.04$. In both cases the width of each histogram bin was adjusted
to be fixed in a logarithmic scale.}
\label{fig3}
\end{figure}

The existence of a truncation in the distribution of sojourn times can be made explicit by
considering cells with smaller sizes, as for example
$\Delta p=0.01$ and $\Delta p=0.04$.  
Figure~\ref{fig3} shows the
counting histogram of sojourn times for two values of $\Delta p$.
The truncation is clearly visible for $\Delta p=0.01$ while a power law tail
is exhibited in the inset for $\Delta p=0.04$. Although different types of truncation
are considered in the literature~\cite{mantegna,gupta}, the most relevant feature for the present study is the existence of
a truncation and where in the distribution it occurs.

In order to obtain an estimate of the crossover from the L\'evy to Gaussian regimes
we use the approach developed in reference~\cite{mantegna}, where the authors consider 
distributions of L\'evy that are symmetric, although, in this problem  
we have one sided asymmetric distributions. Consequently, some adaptations 
to apply that approach must be followed.

We start defining a symmetric frequency distribution $g_{sim}(t)$ obtained from 
the sojourn distribution $g(t)$:
\begin{equation}
\label{ann1}
g_{sim}(t)=\left\{
\begin{array}{ll}
g(-t)/2 & t<0; \\
& \\
g(t)/2 & t>0.
\end{array}\right.
\end{equation}
We approximate $g_{sim}(t)$ by a symmetric truncated L\'evy distribution: 
\begin{equation}
\label{ann2}
\tilde f^{(\alpha)}_{sim}(t)=\left\{
\begin{array}{ll}
\displaystyle cf_{sim}^{(\alpha)}(t) & -L<t<L; \\
& \\
\displaystyle 0 & \mbox{otherwise},
\end{array}\right.
\end{equation}
where $c$ is a constant of renormalization and $f_{sim}^{(\alpha)}(t)$ is a Symmetric Stable L\'evy distribution
with characteristic function given by:
\begin{equation}
\label{ann3}
\psi_{sim}(z)=\exp\left(-\gamma|z|^{\alpha}\right)
\end{equation} 
In order to show that the symmetrized sojourn time statistics for larger $\Delta p$ is well described by 
symmetric truncated L\'evy distribution we perform a sum $X_{sim}^{(n)}$
of $n$ sojourn times randomly chosen among those obtained from our simulations
but also randomly choosing its sign (this is equivalent to symmetrizing the distribution).
Considering $g_{sim}(t)\approx f_{sim}^{(\alpha)}(t)$, then
the probability density for small $n$ is given by~\cite{mantegna}:
\begin{equation}
P^{(n)}(0)=\frac{\Gamma(1/\alpha)}{{\pi\alpha\:(\gamma n)^{1/\alpha}}},
\label{eq3c}
\end{equation}
For $n$ large one has a Gaussian distribution:
\begin{equation}
P^{(n)}(0)=\frac{1}{\sigma\sqrt{2\pi\:n}},
\label{eq3d}
\end{equation}
$\sigma$ being the standard deviation of the symmetrized sojourn distribution $g_{sim}(t)$. 

The crossover $n^{\star}$ from a pure L\'evy to the Gaussian behavior is obtained 
equating equations (\ref{eq3c}) and (\ref{eq3d}); and is given by:
\begin{equation}
\label{ann4}
\displaystyle n^\star=A\sigma^{{2\alpha}/{(\alpha-2)}};\;\;
A=\left[\sqrt{\frac{2}{\pi}}\frac{\Gamma\left(1/\alpha\right)}
{\alpha\gamma^{1/\alpha}}\right]^{{2\alpha}/{(\alpha-2)}}.
\end{equation}
The statistical moments of the distributions $g_{sim}(t)$ and $g(t)$ are related as follows:
\begin{equation}
\label{ann5}
\left<|t|^n\right>_{sim}=\int_{-\infty}^{\infty}|t|^ng_{sim}(t)dt=\int_0^{\infty}t^ng(t)dt=\left<t^n\right>,
\end{equation} 
and, also we have $\left<t^n\right>_{sim}=0$ for all $n$ odd. Thus, the standard deviation $\sigma=\sqrt{\left<t^2\right>}$
can be obtained from the second order moment of the sojourn distribution $g(t)$.

\begin{figure}[ptb]
\begin{center}
\scalebox{0.7}{\includegraphics{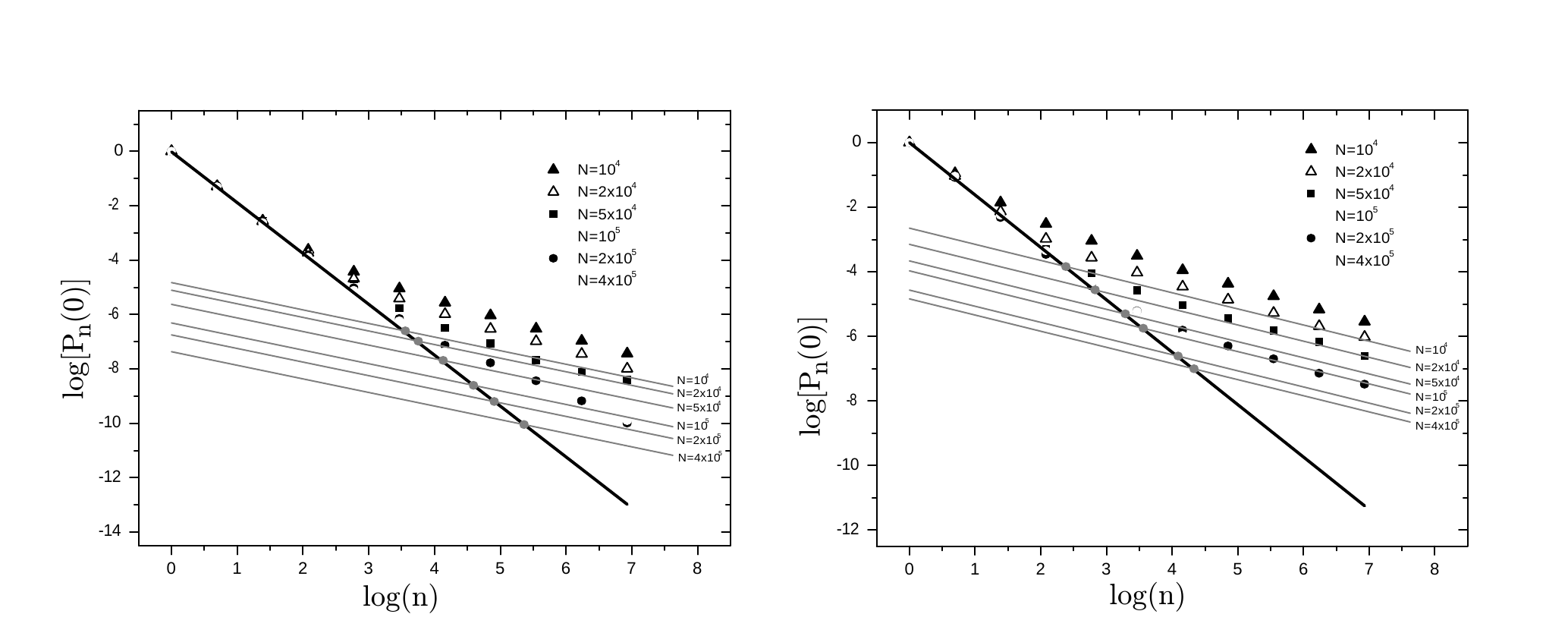}}
\end{center}
\vspace*{-7mm}
\caption{Left Panel: Probability density $P_N^{(n)}(0)$ as a function of the number $n$ of summed
variables for different values of  $N$ and normalized with respect to the probability for $n=1$
for the equilibrium state and averaged over 20 realizations.
The continuous straight line is the curve obtained from a fitting for the initial part of the curve for $N=400,000$
with $\alpha=0.53$ and $\gamma=1.46$.
Right Panel: The same as the left panel but for the waterbag state, the continuous straight line corresponds to $\alpha=0.62$ and $\gamma=2.53$.
The dashed lines is the curve of $P^{(n)}(0)$ given in Eq. (\ref{eq3d}), which corresponds to the Gaussian regime.
The intersection of these lines with the continuous line (the crossover) is marked with gray circles}
\label{fig6}
\end{figure}

Figure~\ref{fig6} shows the probability $P^{(n)}(0)$ as a function of $n$. It strongly suggests that,
regardless of the number of particles,
the random sum process is attracted to a region close to the same symmetric stable L\'evy distribution, which is
well characterized by a given value of $\alpha$ and $\gamma$. We have obtained an estimate of those values by fitting
the first four points of the curve corresponding to $N=400,000$. We remark that the fitted values for the shape exponent $\alpha$
are very closed to the values showed in table~\ref{tab1}, which reinforces the consistency of the methodology based on the 
symmetrization of a one sided distribution.
The most fundamental reason about the consistency between the two approaches is based on the fact that 
the asymmetric and symmetric sojourn distributions have the same power law in their tails.

\begin{figure}[ptb]
\begin{center}
\scalebox{0.3}{\includegraphics{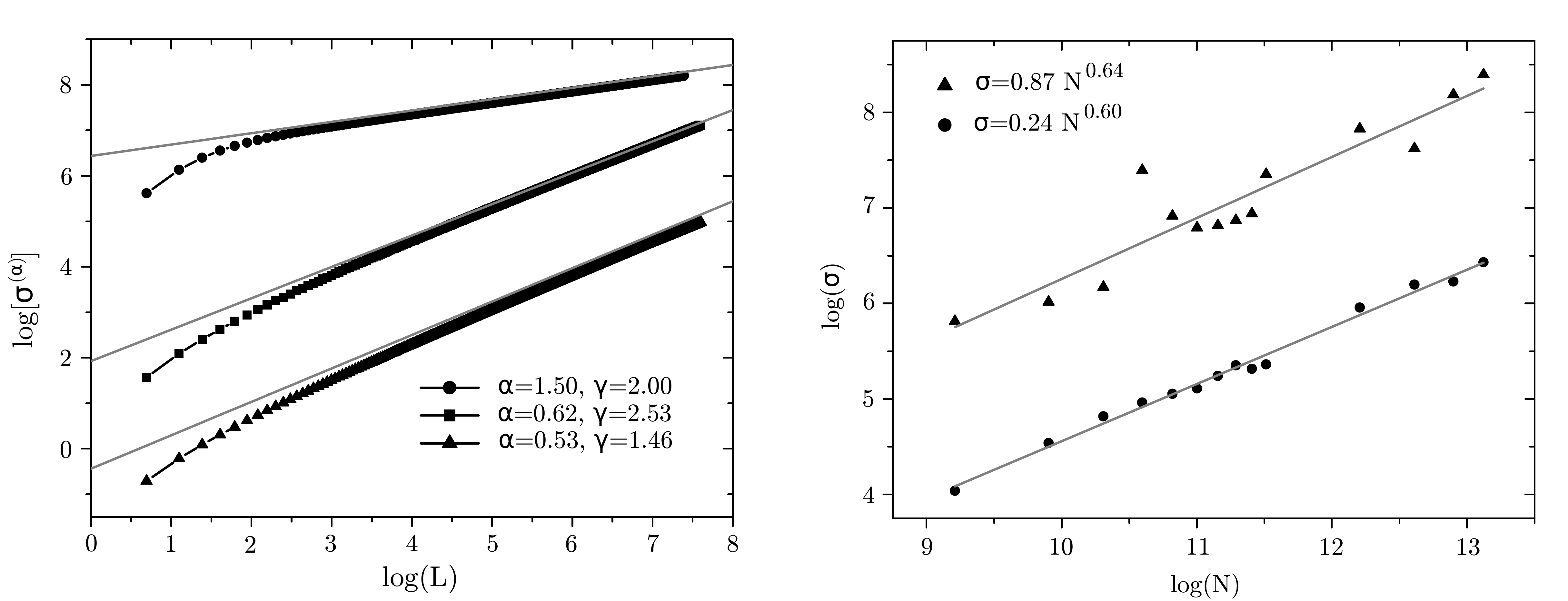}}
\end{center}
\vspace*{-7mm}
\caption{Left Panel: The truncated L\'evy standard deviation $\sigma^{(\alpha)}$ as a function of truncature $L$ for values of 
$\alpha$ and $\gamma$ obtained in
Fig.~\ref{fig6}. The analogous curve for $\alpha=1.5$ and $\gamma=2.0$ is shown for comparison. 
The curves were translated vertically for easier viewing. The gray lines correspond to the asymptotic value given in Eq. (\ref{ann9}).
Right Panel: The empirical standard deviation of sojourn times $\sigma=\langle t^2\rangle$ as a function of the number $N$ of particles.
Triangles correspond to the equilibrium and circles to waterbag states. The best fits for a power-law dependence of $\sigma$ on $N$ are indicated in the
graphics for each case.}
\label{fig7}
\end{figure}

Also, we clearly see from Figure~\ref{fig6} the crossover growth with the number of particles $N$, that is,
the sum variable $X_{sim}^{(n)}$ stays longer times close to a stable L\'evy distribution 
when the number of particles increases. This statement is confirmed by observing the right panel of Figure~\ref{fig7},
where the value of $\sigma$ increases with the number of particles. All the remarks made above are true for both
conditions used in simulations, {\it i.e.}, equilibrium and waterbag.

A relation between $\left<t^2\right>$ and the truncature $L$ in equation (\ref{ann2}) can be obtained (at least asymptotically for $L>>1$) 
through the application of a 
very old result obtained by the french mathematician Paul L\'evy in reference~\cite{levy}. In this work he has showed that if 
the principal value of a given characteristic 
function, associated to a symmetric density distribution $f(t)$, is given by $-\gamma|z|^\alpha$, then
\begin{equation}
\label{ann6}
\lim_{t\rightarrow\infty}f(t)=\frac{B}{t^{1+\alpha}};\;\;
B=\frac{\gamma\Gamma\left(1+\alpha\right)\sin\left(\pi\alpha/2\right)}{\pi};
\end{equation} 
for any $\alpha>0$ with $\sin\left(\pi\alpha/2\right)\neq 0$. This result allows us to calculate the second moment of the
truncated distribution given in (\ref{ann2}):
\begin{equation}
\label{ann7}
\left<t^2\right>_{tr}=c\int_{-L}^{L}t^2f_{sim}^{(\alpha)}(t)dt;\;\;c=\frac{1}{\displaystyle \int_{-L}^Lf_{sim}^{(\alpha)}(t)dt}.
\end{equation}
If $L>>1$ we have $c\approx 1$ and we get the following approximation for Eq. (\ref{ann7}):
\begin{equation}
\label{ann8}
\left<t^2\right>_{tr}\approx 2\int_0^Lt^2f_{sim}^{(\alpha)}dt\approx 
\frac{2B}{2-\alpha}L^{2-\alpha}.
\end{equation}
Where the last equation in the right hand side was obtained using the result in Eq. (\ref{ann6}).
Finally, the standard deviation $\sigma^{(\alpha)}$ of the truncated L\'evy (\ref{ann2}) is given by:
\begin{equation}
\label{ann9}
\sigma^{(\alpha)}=\sqrt{\left<t^2\right>_{tr}}=\sqrt{\frac{2\gamma\Gamma(1+\alpha)\sin\left(\pi\alpha/2\right)}{(2-\alpha)\pi})}L^{(2-\alpha)/2}.
\end{equation}
The calculation of the others statistical moments is straightforward and follows the same steps to obtain Eq. (\ref{ann9}).

\begin{table}[ptb]
\begin{center}
\begin{tabular}{c | c}
\hline\hline
Equilibrium & Water-bag \\
\hline\hline
$\sigma=0.64L^{0.735}$ & $\sigma=0.93L^{0.690}$ \\
$L=1.50N^{0.870}$ & $L=0.14N^{0.869}$\\
\hline
\end{tabular}
\end{center}
\caption{The relationship of $\sigma$ with the truncature $L$ and $L$ with number of particles $N$, respectively for the 
Equilibrium and Water-bag conditions~\cite{trunc}.
These relations are obtained using Eq. (\ref{ann9}) and the fitted power law in the left panel of Fig~\ref{fig7}}
\label{tab2}
\end{table}
 
Figure 6 (left panel) shows the curve of $\sigma^{(\alpha)}$ as a function of $L$ for values of the shape exponent $\alpha$ and
scale factor $\gamma$ obtained by fitting in Figure~\ref{fig6} for both initial conditions: equilibrium and waterbag. 
From this figure we can conclude that for $L>50$ the Eq. (\ref{ann9}) is already a good approximation. 
The divergence of the truncation $L$ can be shown by computing the second order moment $\langle t^2\rangle$ of the empirical sojourn times
for different values of $N$ as shown in the right panel of Figure~\ref{fig7}, where a power law increase with
$N$ becomes evident. This power law combined with Eq. (\ref{ann9}) leads to another power law linking
the truncature $L$ with the number of particles $N$.
In table~\ref{tab2} are shown the respective relations $\sigma(L)$ and $L(N)$.
We remark that the truncature $L$ dependence on the number of particles $N$ seems to be given by the same 
power law for both conditions analyzed.  

From our computational results we can assert that the HMF model becomes ergodic only after a very long time such that the sum of sojourn times
with a truncated L\'evy distribution converges very slowly to a Gaussian distribution~\cite{mantegna}.
This sum yields the distribution of residence times and also determines the ergodic properties of the system.
The time required to attain ergodicity corresponds to the crossover between L\'evy and Gaussian behaviors in
Figure~\ref{fig6}. For shorter times, which are effectively still very long, the system is weakly non-ergodic
and tends asymptotically to the ergodic behavior for finite $N$.
Our results also show that this crossover time diverges with the number of particles in the system.
The results obtained in the present work are in agreement with those presented in Ref.~\cite{eplnos}
where it is shown that the dispersion of the time average of the momentum of a single
particle is non-negligible and becomes small only after a very long time of the order of the relaxation time
to equilibrium, implying that time averages are equal to ensemble averages and thus leading to $F^{(\alpha)}\rightarrow\delta(\overline{\cal O}-\langle{\cal O}\rangle)$.
Up to the authors knowledge this is the first time non-ergodic behavior in a system with long-range
interactions is shown to be related to L\'evy flights in momentum space,
however a more thorough analysis for the HMF model and other models with long-range interactions is under
course and will be the subject of a future publication.

\newpage

A.~F., T.~M.~R.~F.\ and M.~A.~A.\ would like to thank CNPq and CAPES (Brazilian government agencies) for partial financial support and
Z.~T.~O.~Jr.~would like to thank FAPESB for partial financial support. A.~F.\ and T.~M.~R.~F.\
would like to thank R.~Venegeroles for many useful discussions.
Finally, the authors would like to acknowledge the useful comments and suggestions of the referees.


\begin{thebibliography}{}
\bibitem{trunc} The results presented in Table II show the behaviour of $L$ with $N$ adjusted by a 
L\'evy truncated distribution. However, we should point out that the behaviour of $L$ will be discussed thoroughfully in a 
further communication. In this paper we are showing conclusively 
that the sojourn time can be interpreted in the context of L\'evy truncated distributions.
\end{thebibliography}
\end{document}